\documentclass[preprint,aps,prb]{revtex4}

\usepackage{amsmath,amsfonts,epsfig,graphicx,float,color}

\begin{document}

\title{Growing Phenotype-controlled Phononic Materials from Plant Cells Scaffolds}

\author{Maroun Abi Ghanem$^{1\ast}$, Liliane Khoryati$^{2}$, Reza Behrou$^{1}$, Amey Khanolkar$^{1}$, Samuel Raetz$^{3}$, Florian Allein$^{1}$, Nicholas Boechler$^{1}$ and Thomas Dehoux$^{4\dagger}$}

\affiliation{ 
$^1$University of California San Diego, Department of Mechanical and Aerospace Engineering, La Jolla, CA 92093, USA.} 
\affiliation{$^2$Benaroya Research Institute at Virginia Mason, 1201 Ninth Avenue, Seattle, WA 98101, USA.  }
\affiliation{$^3$LAUM, Le Mans Universit{\'e} , UMR CNRS 6613, Avenue Olivier Messiaen, 72085 Le Mans Cedex 9, France}
\affiliation{$^4$Universit\'{e} Claude Bernard Lyon 1, CNRS, Institut Lumi\`{e}re Mati\`{e}re, F-69622 Villeurbanne, France.}
\affiliation{$^\ast$To whom correspondence should be addressed; E-mail:  mabighanem@eng.ucsd.edu.}
\affiliation{$^\dagger$To whom correspondence should be addressed; E-mail:  thomas.dehoux@univ-lyon1.fr.}

%\begin{affiliations}
%\item University of California San Diego, Department of Mechanical and Aerospace Engineering, La Jolla, CA 92093, USA
%\item Benaroya Research Institute at Virginia Mason, 1201 Ninth Avenue,Seattle, WA 98101, USA
%\item Laboratoire d'Acoustique de l'Universit{\'e} du Mans, LAUM - UMR 6613 CNRS, Le Mans Universit{\'e}, Avenue Olivier Messiaen, 72085 Le Mans Cedex 9, France
%\item Universit\'{e} Claude Bernard Lyon 1, CNRS, Institut Lumi\`{e}re Mati\`{e}re, F-69622 Villeurbanne, France
%\end{affiliations}

\maketitle

\textbf{
Biological composites offer self-healing properties, biocompatibility, high responsivity to external stimuli, and multifunctionality, due, in part, to their complex, hierarchical microstructure. Such materials can be inexpensively grown, and self-assembled from the bottom up, enabling democratized, sustainable manufacturing routes for micro- and nano-devices. While biological composites have been shown to incorporate rich photonic structures, their phononic properties have hitherto remained unexplored. In this study, we demonstrate that biological composites in the form of micron-thick decellularized onion cell scaffolds behave as an organic phononic material, with the presence of band gaps forbidding the propagation of elastic waves in select frequency ranges. We show that the onion cells' phononic properties can be phenotypically tuned, and anticipate these findings will yield new biologically-derived, ``green," and genetically tailorable phononic materials.
}

From a sustainability perspective, it is critical that existing technologies incorporating inorganic compounds, such as rare earth metals, be replaced with biodegradable and recyclable organic materials.\cite{mohanty2018composites} Because of this, the development of biological composites has become a societal necessity.\cite{liu2017functional, eder2018biological} As an added value, biological composites also offer self-healing properties,\cite{liu2016water} biocompatibility,\cite{gershlak2017crossing} high responsivity to external stimuli for applications such as sensing and soft robotics,\cite{DiGiacomoeaai9251, fratzl2009biomaterial} and multifunctionality (for instance, as is exemplified by butterfly wings having both structural coloration\cite{vukusic2003photonic} and superhydrophibicity\cite{darmanin2015superhydrophobic}). In contrast to inorganic systems, which are typically fabricated using non-scalable top-down approaches (where the fabrication time scales cubically with the system to microstructure size ratio), biological composites exhibiting complex structural hierarchy can be grown, and scalably self-assembled from the bottom up.\cite{wegst2015bioinspired, barthelat2016structure} Such naturally-occurring structural hierarchy makes biological composites a class of uniquely transformative materials for bridging functional nanocomposites to macroscale integrated devices in a green manner.\cite{begley2019bridging} 

Within the context of biological composites, a wide array of systems with rich functionalities have been implemented, such as cellulose-based flexible electronics,\cite{zhang2018flexible} plasmonic wood,\cite{jiang2018wood} wool-based nano-patterns,\cite{zhu2019using} lasers,\cite{nizamoglu2013all} and DNA-based mechanical metamaterials.\cite{lee2012mechanical} Yet harnessing the power of biological composites to control elastic waves has remained an unmet challenge, partly because, in contrast with visible photons, there has been no observation of phonon-based biological function at the supramolecular scale. In the past decade, the use of inorganic, man-made composites, often referred to as phononic materials,\cite{hussein2014dynamics} have enabled manipulation of elastic waves across frequencies ranging from tens of Hz to a few THz, with applications ranging from seismology to thermal transport, respectively.\cite{maldovan2013sound} As a result of their structural features, inorganic phononic materials have been shown to exhibit unique properties such as negative Young's modulus, mass density, and refraction.\cite{ma2016acoustic} Integration of these materials in ultrasonic biosensors has provided sensitivity enhancements\cite{Xu2018_v99_p500--512, Lin2019_v24_p2722--2727} and enabled diagnostic applications.\cite{Reboud2012_v_p} Miniaturized lab-on-a-chip devices for fluid shaping,\cite{Bourquin2019_v23_p1458--1462, Wilson2011_v11_p323-328} nebulization,\cite{Reboud2012_v12_p1268-1273} tweezing,\cite{Ozcelik2018_v15_p1021--1028} and streaming and sorting\cite{Hashmi2012_v12_p4216-4227} have all shown benefit from the introduction of phononic microstructures. The use of phononic materials in ultrasonic signal processing, such as the SAW filters which are ubiquitous in modern communication devices,\cite{Xu2018_v99_p500--512} has been shown to reduce device size and enable functionalities such as wave guiding and logic.\cite{Yan2018_v17_p993--998}

However, these inorganic phononic materials operating at high frequencies (MHz and above) are subject to several major limitations, namely manufacturing scalability,\cite{Cummer2016_v1_p16001} fragility, sustainability, and biocompatibility. Leveraging the ability of biological composites to be grown rapidly, inexpensively, en masse from renewable resources would allow building biodegradable and recyclable phononic materials, and facilitate the democratization of state-of-the-art technologies in the areas of sensing, microfluidic control, acoustic signal processing, and potentially even thermal control materials. Moreover, in contrast to their inorganic counterparts, biological materials are soft, stretchable, and inherently biocompatible, making them uniquely suited to translate the above-mentioned applications into wearable sensors,\cite{Jin2013_v3_p2140, Li2017_v8_p15310} health-care devices,\cite{Kang2018_v4_p} and human augmentation devices.\cite{Lee2019_v24_p6914--6921} Indeed, the discovery of piezoelectricity in plants\cite{ciofani2012piezoelectric} suggests the possibility to create purely biological SAW devices with phononic material enhanced functionalities. We also envision these biocompatible phononic materials could used in ultrasonic biomedical imaging\cite{Hu2018_v4_p} and wave-assisted regenerative medicine applications\cite{tsimbouri2017stimulation} as integrated signal focusing and localization devices\cite{Jin2019_v10_p143} or acoustically-encoded reporters.\cite{bourdeau2018acoustic} We note that a prior claim of phononic band gap observation\cite{schneider2016nonlinear}  in a biological composite was later found to be erroneous\cite{wang2020determination}. In this paper, we demonstrate that biological composites in the form of decellularized plant cell scaffolds can behave as phononic materials, including forbidding the propagation of elastic waves in select frequency ranges (i.e. band gaps). Our discovery of phononic behavior in biological composites, in particular, plants, shows how biology can be directly used to transfer the existing physics of phononic materials into scalably manufacturable, biocompatible, sustainable, and democratizable forms. 

Our biological composite is composed of a micron-thick onion cell epidermis with slender cell walls extruding up from it that resemble blind bore beehive structures, and is adhered to a glass substrate. We measure the dispersion of sub-GHz surface acoustic waves (SAWs) in the onion composite, and reveal their interaction with compressional and flexural resonances of the wall structure, which open deep and wide band gaps. We also demonstrate that the SAW dispersion can be phenotypically controlled by selecting plants at different developmental stages. We thus foresee that, in the long-term, synthetic biology, whereby artificial devices are built from biological parts (cells, DNA strands...),\cite{smanski2016synthetic} and the ability to tailor the genome by controlled mutations or gene editing,\cite{belhaj2013plant} could provide a plausible, scalable manufacturing route for future phononic materials design. 
%=========================
%== Results and Discussion ===
%=========================

\begin{figure}
\begin{center}
\includegraphics[width=17cm]{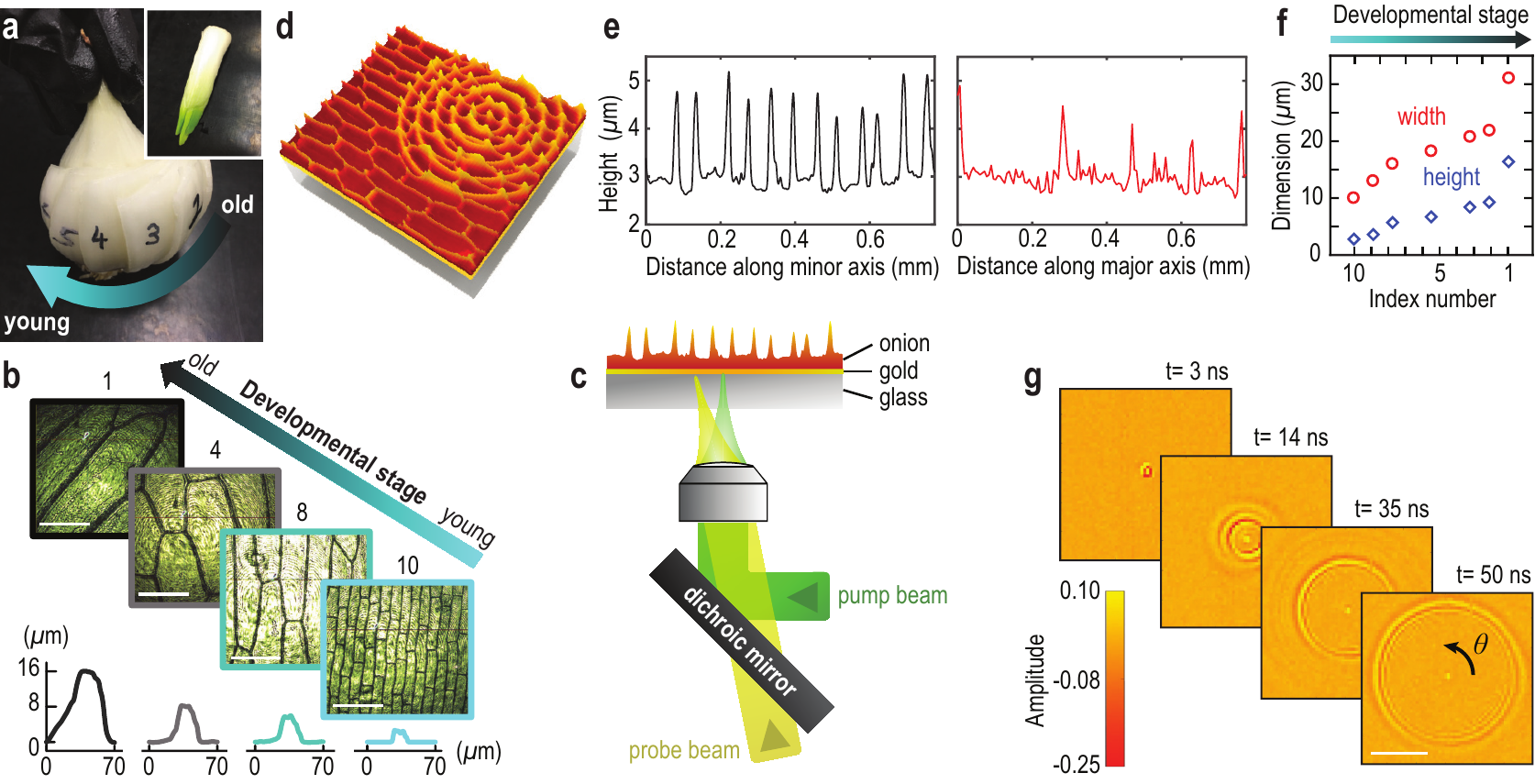}
\end{center}
\caption{\label{sample} (a) Sample preparation. Epidermal layers were obtained from different depths of an onion bulb, going from the outermost (oldest) to the innermost scale (youngest), shown in the inset. (b) Optical microscope images of the peeled organic layers showing different morphogenetic profiles as a function of their developmental stage. Scale bar is 100 $\mu$m. The diagrams show profiles of the pillar-like structure of the wall as a function of the developmental stage (index denotes layer number). (c) Opto-acoustic setup used to generate and detect sub-GHz SAWs in the sample. (d) Profilometry image of the onion epidermal surface with a superimposed illustration of the propagating SAWs. (e) Representative profiles along the minor and major axis of the cells, respectively. (f) Variation of the height and width of the wall's vertical portions as a function of the developmental stage of the plant.(g) Snapshots at different times showing the propagation of SAWs on a bare substrate. Scale bar is 100 $\mu$m.}
\end{figure} 

Fresh onion scales were selected from different depths of an onion bulb, as is shown in \textbf{Fig.~\ref{sample}a}. The outer half of selected layers was peeled off and dehydrated over 1-2 days at 4$^\circ$C prior to testing (\textbf{Supplementary note 1}). \textbf{Figure~\ref{sample}b} shows optical microscope images of the peeled organic layers, which show different phenotypes corresponding to different developmental stages (indexed as a function of consecutive layer number, starting from the outermost epidermis). The scales were placed onto a photoacoustic transducer composed of a 100 nm layer of gold coated onto a thick 1~mm glass substrate, as is illustrated in \textbf{Fig.~\ref{sample}c}. The remaining cell walls, constituting the bottom of the onion scale and the vertically protruding walls, resemble a borehole beehive structure. We show a 3D profilometry image of a representative sample surface in \textbf{Fig.~\ref{sample}d}, as well as typical line profiles recorded along the minor and major axes of the cavities in \textbf{Fig.~\ref{sample}e} (which corresponds to the sample shown in \textbf{Fig.~\ref{sample}d}). The surface topography of the decellularized epidermal layers reveals that the vertical portions of the cell walls decrease in size (height and full width at half maximum) as a function of the age of the layer, ranging from about $16$~$\mu$m~$\times~30$~$\mu$m for the oldest layer to $3$~$\mu$m~$\times10~\mu$m for the youngest layer, as is shown in \textbf{Fig.~\ref{sample}f}.

To analyze the sub-GHz acoustic properties of this organic structure, we illuminate the bottom of the metal film with a pulsed laser (400~ps pulse duration, 532 nm wavelength, and 1.0 mW average power) focused to a $\sim10$ $\mu$m diameter spot (\textbf{Fig.~\ref{sample}c} and \textbf{Supplementary note 2}). The subsequent rapid thermoelastic expansion in the metal generates propagating SAWs with a broad frequency spectrum extending up to 400~MHz. We use the time-dependent deflection of a continuous probe beam (577 nm wavelength, 12 mW average power) by the acoustic-induced surface ripples (sketched in \textbf{Fig.~\ref{sample}d}) to monitor the propagation of the SAWs (\textbf{Supplementary note 2}). By scanning the probe beam across the bottom of the metal film over a $300\times300$~$\mu$m$^2$ area, we obtain movies of the propagating SAWs. These animated maps reveal largely circular wavefronts, as is shown in \textbf{Fig. 1g}, which depicts SAWs propagating along a substrate without a cell layer on top (hereafter referred to as ``bare'').

\begin{figure}
\begin{center}
\includegraphics[width=138mm]{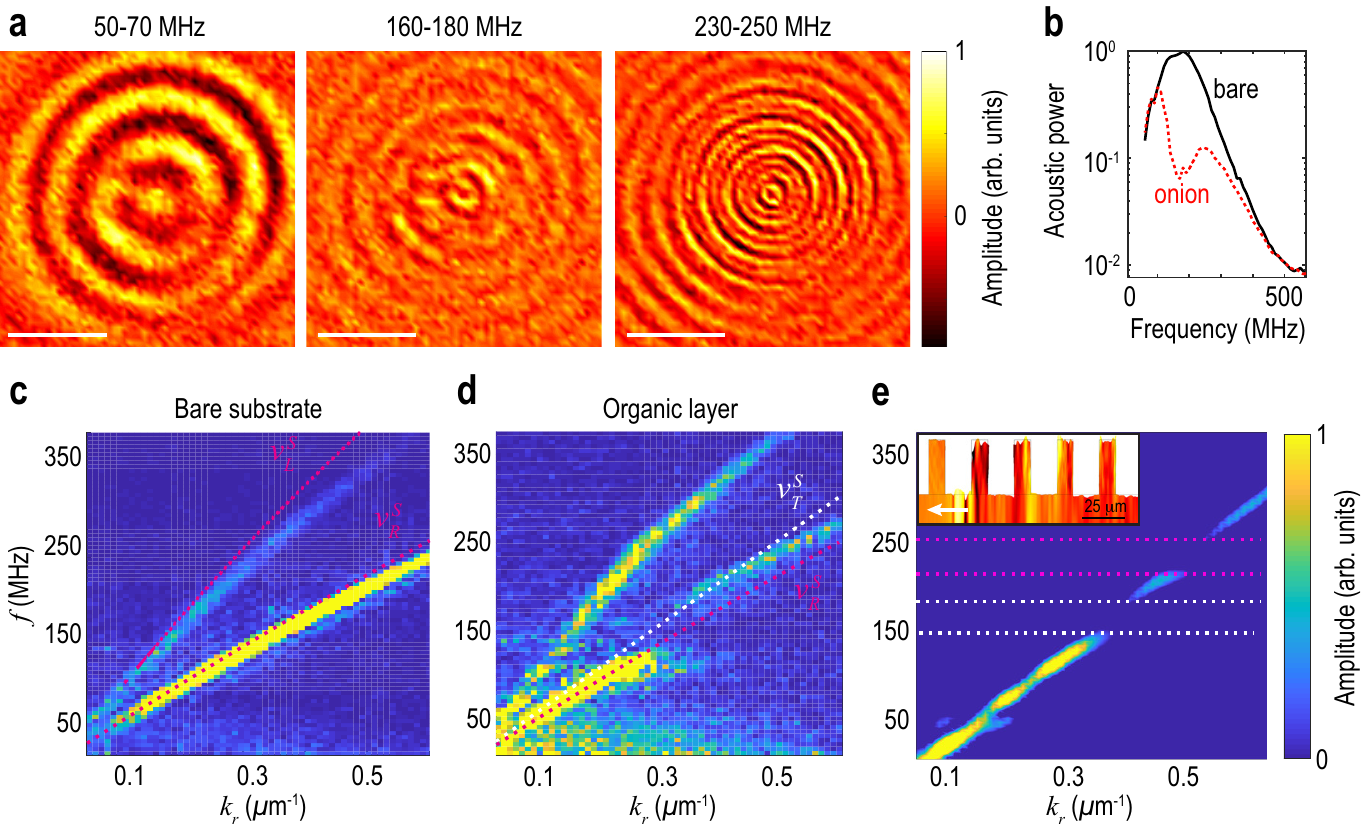}
\end{center}
\caption{\label{Fig2} (a) Snapshots at $t=40$ ns, filtered to display wave propagation in limited frequency ranges. Scale bar is 100 $\mu$m. (b) Acoustic power calculated at each frequency for the bare (black) and for the onion epidermis (red). Dispersion curve measured on (c) the bare substrate and (d) the organic layer. The magenta dotted lines in (c) correspond to the dispersion of surface and longitudinal wave speeds calculated using a layered-half-space model. The white dashed line corresponds to the transverse wave speed in the bare substrate. (e) Dispersion curve obtained from the numerical analysis. The inset shows the geometry used in the numerical model and the simulated SAW in the time domain is indicated by white arrows. The snapshot is taken at $t$ = 35 ns.}
\end{figure}

During its propagation, the SAW sets the onion epidermis atop in motion, and this coupled movement alters the SAW dispersion. To show this, we plot snapshots of the traveling SAWs in \textbf{Fig. 2a}, filtered to display wave propagation in three limited frequency ranges (\textbf{Supplementary note 3}). We observe very low acoustic amplitude $u_i^f(x,y)$ in the 160-180 MHz range, compared to the other two frequency ranges. Such frequency dependence is not present for SAWs propagating across the bare substrate. In \textbf{Fig.~\ref{Fig2}b}, we plot the acoustic power $P_i^f=\sum_{x,y}|u_i^f(x,y)|^2$ calculated at each frequency $f_i$, for the bare substrate (black) and for the onion epidermis characterized in \textbf{Fig.~\ref{Fig2}a} (dashed red line) at a time $t\sim50$~ns. This plot indicates the presence of a phonon stop band, or resonant attenuation zone, in the 160-180 MHz range with 94\% extinction compared to the bare substrate caused by the presence of the organic layer.

To better understand this observation, we calculate the acoustic dispersion curves $|\tilde{u}(k_r,\theta,f)|$ using a 2D Fast Fourier Transform (FFT) in the radial wavevector-frequency domain $k_r$-$f$ for each angular direction $\theta$ (see \textbf{Fig.~\ref{sample}g} and \textbf{Supplementary note 3}). \textbf{Figure~\ref{Fig2}c} shows a dispersion curve measured on the bare substrate for one wavevector direction. We observe two distinct modes, corresponding to surface skimming longitudinal and Rayleigh SAW modes. We calculate analytical dispersion curves using a slow-on-fast layered-half-space model (dotted red lines in \textbf{Fig.~\ref{Fig2}c}),\cite{Ewing1957_v_p} with typical acoustic properties for silica ($v_L^s=5700$~m/s, $v_T^s=3400$~m/s, and $\rho^s=2500$~kg/m$^3$ \cite{Bansat1986_v_p} ) and gold ($v_L^g=3200$~m/s, $v_T^g=1200$~m/s, and $\rho^g=19320$~kg/m$^3$ \cite{Kutz2013_v_p}), and find good agreement with the experimentally identified branches. 

The dispersion curve (\textbf{Fig.~\ref{Fig2}d}) corresponding to a location containing an epidermal layer shows a starkly different behavior. We observe avoided crossing behavior due to the coupling of Rayleigh waves with local resonances in the organic layer, that opens a wide gap (likely aided by resonant attenuation effects) around 170 MHz. The lower branch starts as a Rayleigh wave at low-$k_r$ values (red dotted line) and approaches a horizontal asymptote near the resonance frequency. The upper branch tends to the Rayleigh wave speed at high wavenumbers, but deviates from the Rayleigh line near the gap frequency and vanishes beyond the $v_T^s$ threshold (white dotted line), below which it becomes evanescent.\cite{Every2002_v13_pR21--R39} Such dispersion curves are characteristic of locally resonant metamaterials, and have previously been observed at the geophysics scale,\cite{Colombi2016_v6_p19238} wherein trees in a forest were used as local resonators, and down to the microscale via the interaction of SAWs with the contact resonances of microsphere monolayers.\cite{Hiraiwa2016_v116_p198001}

\begin{figure}[h]
\begin{center}
\includegraphics[width=127mm]{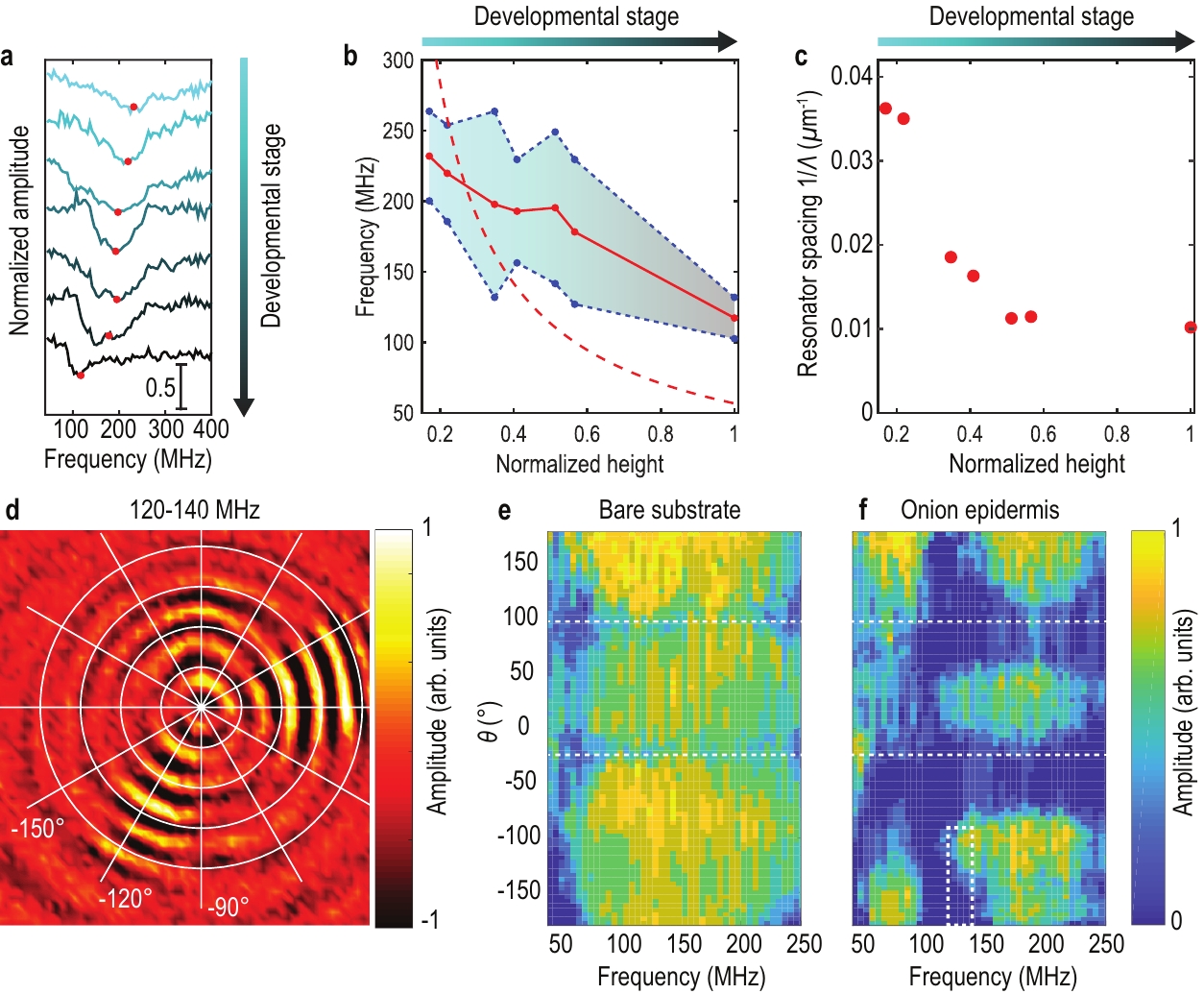}
\end{center}
\caption{\label{3} (a) Spatial average of the frequency response normalized by the response of the bare substrate for seven layers of increasing age. Red dots indicate the center of the gap, also reported in \textbf{Fig.~\ref{3}b} (b) Variation of the phenotype-controlled gap as a function of the cell growth stage. The solid red line indicates the center frequency of the band gap. The dashed red line shows the variation of the first compressional mode (fixed-free organ-pipe mode) of a continuous onion pillar as a function of the normalized height. The blue shaded region indicates the edges of the gap, defined as 70\% of the average amplitude level outside of the gap. (c) Variation of the resonator spacing with the developmental stage of the cells. (d) Snapshot at $t$ = 60 ns taken from an acoustic movie filtered between 120-140 MHz, that shows directionality in the SAW propagation. $f-\theta$ maps corresponding to (e) the bare substrate and (f) the onion epidermis.}
\end{figure} 

By analogy with observations made on inorganic, man-made phononic metamaterials,\cite{Hartschuh2005_v87_p173121, Graczykowski2014_v104_p123108--} we suspect the slender vertical portions of the wall may serve as the local resonators that couple to the SAWs. To verify this assumption, we conduct 2D finite element (FE) analysis using COMSOL Multiphysics. We model SAWs propagating on a glass substrate by applying a point-like load with a step-like excitation in the time domain (\textbf{Supplementary note 4}). To mimic the response of the wall's vertical portions, we add rectangular ridges, equally spaced by 26 $\mu$m, with a height of 5 $\mu$m and width of 10 $\mu$m, as is illustrated in the inset of \textbf{Fig. 2e}. The ridges are given the effective elasticity of dry cell walls (\textbf{Supplementary note 4}).\cite{Xu2004_v38_p363-374} From our simulated spatiotemporal diagrams, we plot the dispersion curve of the SAWs using a 2D FFT, as is shown in \textbf{Fig. 2e}. Our computational analysis reveals two wide gaps (denoted by white and red dashed lines) between 150 and 250 MHz, which qualitatively supports our observation of band gaps in this frequency range. The complexity of the time-domain modes shapes (see inset of \textbf{Fig. 2e}) suggests that the gaps we observe originate from the hybridization of the propagating SAW with an abundance of closely-spaced multi-resonant modes (both flexural and compressional), similar to what has been observed for Lamb waves in macroscale inorganic locally resonant acoustic metamaterials.\cite{Rupin2014_v112_p234301} We note that, in contrast to typically locally resonant metamaterials where the resonators are uncoupled, because of the inter-resonator connectivity (corresponding to the beehive structure) in the physical sample, we expect the hybrid mode with a horizontally-asymptotic branch induced by local resonances to revert to a propagating mode with a non-zero dispersion curve at high $k$.\cite{Vega-Flick2017_v96_p024303} The multiresonant nature of the ridges could be leveraged to realize locally-resonant metamaterials that have increased tunability with minimal structural complexity, in comparison to phononic crystals or metamaterials having a single local resonance.

The structure of plants, which is one element of their phenotype, can be controlled by environmental or genetic cues. Such modifications, which have enabled crop optimization, are also at the core of plant synthetic biology.\cite{Medford2014_v346_p162--163} To illustrate the ability to control the structure of the organic resonators without any physical or chemical intervention, we select epidermises at different developmental stages. As the plant grows, the phenotype changes and the ridges become thicker and higher and further spaced, as observed in the profilometry images (\textbf{Fig.~\ref{sample}f}). In \textbf{Fig.~\ref{3}a}, we plot the spatial average of the displacement amplitude $\tilde{u}_{av}(f)=\sum_{k_r,\theta}\tilde{u}(k_r,\theta,f)$ calculated at each frequency, normalized by the response of the bare substrate for seven layers of increasing age from the same onion bulb (\textbf{Supplementary note 3}). We observe a gap whose center frequency decreases with the age of the epidermis by up to a factor of two, from $\sim$230 MHz to $\sim$120 MHz (denoted by the solid red line in \textbf{Fig.~\ref{3}b}).

Among the different parameters that can influence the gap tunability, let us first discuss the influence of the geometry of the cell wall (resonators) and consider a simple analytical description. Rayleigh waves, having an elliptical polarization, are known to hybridize strongly with compressional resonances.\cite{Hiraiwa2016_v116_p198001} We thus estimate the frequency $f = v_L^L/4h$ of the first out-of-plane compressional mode (organ-pipe mode) of a continuous onion pillar, where $v_L^L = 3400$ m/s is the average longitudinal sound speed measured in onion cells at high frequencies\cite{Gadalla2014_v239_p1129--1137} and $h$ is the height of the ridges measured by profilometry. We plot in \textbf{Fig.~\ref{3}b} the frequency of this mode as a function of normalized ridge height and developmental stage (dotted red line). We observe a qualitative agreement in the frequency range and trend of the measured gaps as a function of ridge height, illustrating that the increase in the height of the vertical portions of the wall, as a result of tissue growth, correlates to a decrease in the gap frequency. Similarly, a decrease in the average stiffness of the epidermis could also result in a decrease in the gap frequency. The plant cell wall is mainly composed of highly oriented cellulose microfibrils cross-linked by hemicelluloses, embedded in a hydrated gel-like matrix of pectins.\cite{Hansen2011_v155_p246--258, Suslov2006_v57_p2183--2192, Davies2003_v217_p283--289} To regulate cell growth, new layers of cellulose are deposited with highly oriented microfibrils, which result in older cell walls being stiffer. We suggest that this change in elasticity may partially counteract the ridge-height-induced decrease in band gap frequency with increasing cell age. 

We also quantify the width of the band gap (whose edges are defined as 70\% of the average amplitude level outside of the gap) and plot it as a function of the developmental stage (indexed as the normalized height of the organic layers's vertical portions), as is shown by the blue shaded area in \textbf{Fig.~\ref{3}b}. We see that the gap becomes narrower as the organic layer ages. To investigate this, we consider the spatial density of the resonators. We estimate the averaged spacing $\Lambda$ between the cells by counting the number of vertical wall portions along the minor axis of the cell on the profilometry images. In \textbf{Fig.~\ref{3}c}, we plot 1/$\Lambda$ as a function of normalized ridge height. We see that $1/\Lambda$ decreases with the age of the organic layers, i.e. the spatial density of the resonators decreases, which is consistent with the observed narrowing of the band gap with increased age. Our observations show that geometry, stiffness and number of resonators are key in adjusting the features ---center frequency and width--- of this phenotype-controlled band gap. These parameters of the biological composite could be manipulated by selecting the developmental stage (as we did), or by mutations or gene editing to tune phonon dispersion precisely, opening sustainable routes to engineer scalably manufacturable metamaterials from grown plants. 

The growth of oriented microfibrils contributes to cell elongation, and thus a transition from isotropic to orthotropic structure. In some of our samples, we observed direction-dependence of the band gap. We plot in \textbf{Fig. 3d} a snapshot of propagating SAWs in the first onion layer (i.e. the oldest), filtered in the 120-140~MHz range. In the lower left quadrant, we see that acoustic energy propagates around -120$^\circ$, while no acoustic energy remains after -150$^\circ$. To explain this, for each angle $\theta$, which denotes the direction of the radial wave vector, we project the $f-k$ dispersion maps on the $f$-axis to obtain the average displacement amplitude as a function of angle, as is shown in \textbf{Figs. 3e,f} (see \textbf{Supplementary note 3}). For the bare substrate (\textbf{Fig. 3e}), we observe two regions with low energy around $100^\circ$ and $-80^\circ$ that are due to a lower sensivity of the laser-deflection probe. This artifact was visible on all the epidermises we probed, as well as on the bare substrate (\textbf{Fig. 3e}). In contrast, in \textbf{Fig. 3f}, which corresponds to the oldest onion layer we probed, we observe a gap extending almost over all the $f-\theta$ map that has a clear $\theta$-dependence. We highlight in \textbf{Fig. 3f} with a white square a region corresponding to the lower left quadrant in \textbf{Fig. 3d} (between -180 and -90$^\circ$). There, we observe that the gap opens and closes successively in the 120-140~MHz range, supporting the observation we made in the filtered image (\textbf{Fig.~\ref{3}d}). While natural anisotropy of the plant epidermis can lead to such directional band gaps, we note that microfibril reorientation can also be forced by repeated loading cycles (making the epidermis stiffer in the direction of the applied strain.\cite{Suslov2009_v60_p4175-4187}) Such conditioning may offer yet another means to control plant-based metamaterials. By leveraging the multiple advantages provided by biological composites, we anticipate our advances will lead to a wide range of new green and sustainable microdevices with tailored functionalities.

\end{document}